\documentstyle[twoside,fleqn,npb,epsfig]{article}
\newcommand{\xgo}{\mbox{$x_{\gamma}^{\rm obs}$}}

\newcommand{\AmS}{{\protect\the\textfont2
  A\kern-.1667em\lower.5ex\hbox{M}\kern-.125emS}}

\title{Semi-leptonic Decays of Heavy Quarks in Dijet \\Photoproduction
       at HERA}

\author{M. Wing \address{McGill University, Montreal, Canada} (On behalf of 
        the ZEUS collaboration)}

\begin{document}

\begin{abstract}
The production of heavy quarks has been studied in photoproduction 
processes with the ZEUS detector at HERA using an integrated luminosity 
of 36.9 ${\rm pb^{-1}}$. Events with a photon virtuality, 
$Q^2<1 \ \rm{GeV^2}$, were selected with two jets of high transverse energy  
and an electron in the final state. Consideration of the distribution in 
$p_T^{\rm rel}$ $-$ the momentum of the electron 
transverse to the axis of the jet to which the electron is closest $-$ allows 
a measurement of the beauty cross-section in a restricted region of phase 
space.
\end{abstract}

\maketitle

\section{Introduction}

Two important motivations for the study of heavy quarks are the 
possibility of providing stringent tests of perturbative QCD and 
yielding knowledge of the structure of the photon. In this paper, the 
emphasis is to make a measurement of beauty production, which, with its 
larger mass should allow perturbative QCD calculations to be more reliable. 
The treatment of heavy quarks is one of the uncertain 
inputs to parametrisations of the photon structure function. Consideration 
of \xgo, where \xgo \ is the fraction of the photon energy contributing 
to the two highest transverse energy jets, may give us some insight into the 
structure of the photon. We define \xgo \ in terms of the transverse 
energy, $E_T^{\rm jet}$, and pseudorapidity, $\eta^{\rm jet}$, of the 
two highest transverse energy jets;

\begin{eqnarray}
 \xgo = \frac{\sum_{jet1,2} E_T^{\rm jet} e^{-\eta^{\rm jet}}}
            {2yE_e},
\end{eqnarray}

\hspace{-0.4cm}where $yE_e$ is the initial photon energy. Results from the production 
of $D^*$(2010) mesons in dijet photoproduction \cite{dstar} show a significant 
cross-section at low \xgo \ consistent with approximately 45\% 
 resolved photon processes on comparison 
with {\sc Herwig} Leading Order (LO) Monte Carlo (MC) \cite{herwig} 
predictions. This demonstrates that at LO, we should consider charm 
production not just as a boson gluon fusion process, but also as a result of 
the photon fluctuating into a source of partons.
% and whether the charm quark is an active parton in the photon 
%$-$ something currently unknown. 
It is then natural to pose the question of  the dominant 
mechanisms in beauty production.

\section{Cross-section Definition}

For the process, 

\begin{eqnarray}
\vspace{-1cm}
e^+ + p \rightarrow {\rm dijets} + e^- + X,
\end{eqnarray}

\hspace{-0.4cm}two differential cross-sections, $d\sigma/d\xgo$ and $d\sigma/dp_T^{\rm rel}$ 
were measured, inclusive of quark flavour. The quantity
\xgo \ is defined in Equation (1) and $p_T^{\rm rel}$ is the momentum of 
the electron transverse to the axis of the jet to which the electron is 
closest. A data sample with a total luminosity of 36.9 ${\rm pb^{-1}}$ 
collected during the years 1996 and 1997 was analysed.

The virtuality of the photon was required to be $Q^2<1 \ {\rm GeV^2}$ which 
allied to $0.2<y<0.8$, where $y$ is the fraction of the the electron's energy 
carried by the photon, defines the events to be in the photoproduction 
region. Each event was required to have two high transverse energy 
jets (reconstructed 
using the $k_T$ clustering algorithm \cite{kt}) with 
$E_T^{\rm~jet1,2}>7,6 {\rm~GeV}$ and $|\eta^{\rm jet}|<2.4$. An 
electron in the final state was selected in the event with 
$p_T(e^-)>1.6 \ {\rm GeV}$ and $|\eta(e^-)|<1.1$.

\section{Method}

The electrons were identified using their rate of energy loss, 
$dE/dx$, traversing the Central Tracking Detector (CTD). 
With the knowledge that for electrons, $dE/dx \sim 1.4$ mips (minimum 
ionising particles) and hadrons, with $p_T(e^-) >$ 1.6 GeV, $dE/dx$ 
$\sim$ 1 mip, a separation of the two can be 
realised. The resolution of $dE/dx$ is not such that direct identification 
can be performed, so we require additional information. 

If a cluster of cells in the calorimeter has more than 90\% of the energy 
deposited in the electromagnetic calorimeter, we call it \emph{electron 
enriched}. If a cluster has more than 60\% of 
the energy deposited in the hadronic calorimeter, we call it \emph{hadron 
enriched}. These clusters are then matched to tracks 
and the $dE/dx$ considered. Although, the electron enriched sample contains 
mostly hadrons, the hadron enriched sample is an efficient rejector of 
electrons. Therefore a statistical subtraction of the two yields the 
number of electrons.

%, the signal being seen as a function of $dE/dx$. 

Before using the electron sample to extract physics results, we  
have to eliminate any remaining background. The significant, background 
arises from photons converting to an $e^+e^-$ pair in dead material.
The pairs are identified with a topological finder, which 
considers the distance of closest approach, vertex and invariant mass of 
all combinations of two tracks. Using a a calculation 
for pair production \cite{tsai}, the number of pairs missed is estimated when 
the $e^+$ has low momentum (below 200 MeV) such that the CTD reconstruction 
efficiency is poor. A second step relies on MC, but at a reduced 
level as a consequence of the model independent first step.

\begin{figure}[htb]
\vspace{-0.5cm}
\begin{center}
~\epsfig{file=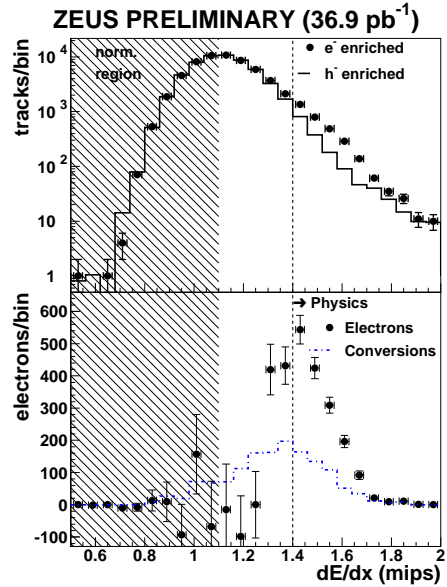,height=8cm}
\end{center}
\vspace{-1.25cm}
\caption{Distribution in $dE/dx$ for electron and hadron enriched samples 
(top) and electron signal with the conversion background (bottom).}
\label{fig:signal}
\vspace{-1.cm}
\end{figure}

In Figure \ref{fig:signal}, we see the $dE/dx$ distribution for electron 
and hadron enriched samples (top) and the electron signal with the 
background from converting photons (bottom). The hadron enriched sample 
is normalised to the electron enriched sample in the shaded region shown. 
In this region a good description of the hadronic background in the electron 
enriched sample by the hadronic enriched sample is seen. An excess of the 
electron enriched sample over the hadron enriched sample at larger values 
of $dE/dx$ is also seen, consistent with the presence of electrons, which is 
then clearly shown in the subtracted plot. In Figure~\ref{fig:signal} (bottom)
a clear electron signal is seen which allied to the background estimation 
can then be used to extract physics results.

\section{Results}

\begin{figure}[htb]
\vspace{-1.2cm}
\begin{center}
~\epsfig{file=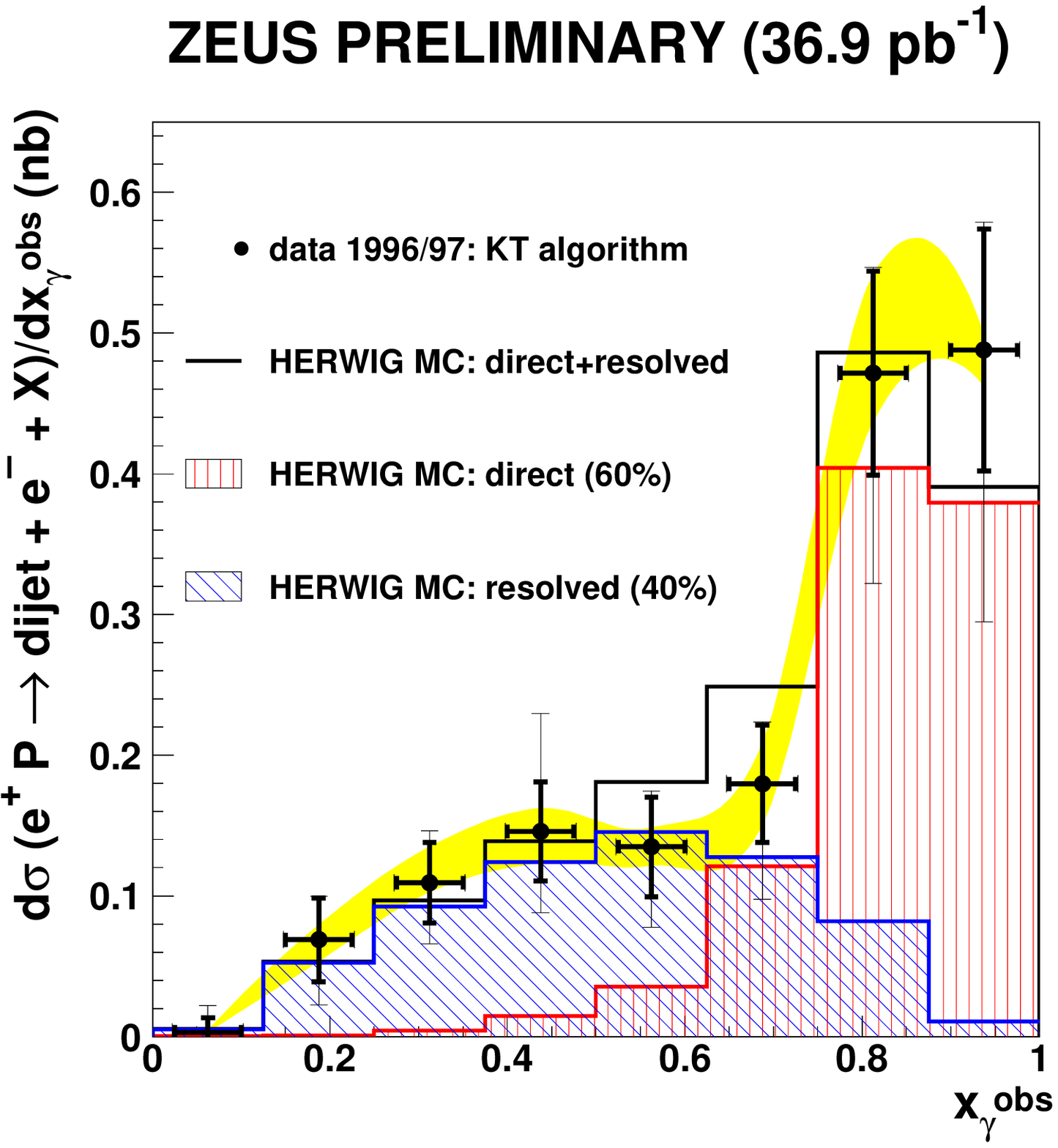,height=7cm}
\end{center}
\vspace{-1.35cm}
\caption{Cross-section $d\sigma/d\xgo$ compared to MC 
expectations.}
\label{fig:xgamma}
\vspace{-0.7cm}
\end{figure}

\begin{figure}[htb]
\vspace{-1.2cm}
\begin{center}
~\epsfig{file=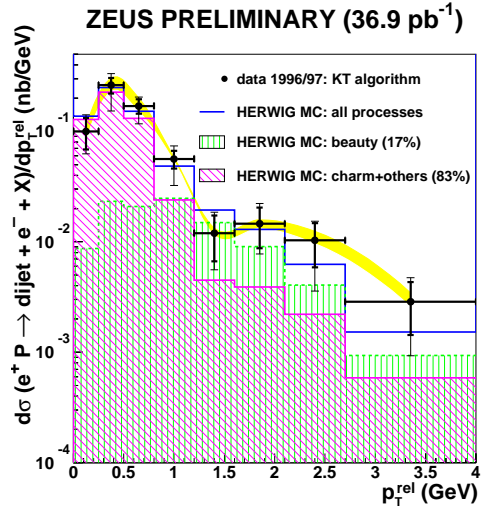,height=7cm}
\end{center}
\vspace{-1.3cm}
\caption{Cross-section $d\sigma/dp_T^{\rm rel}$ compared to MC 
expectations.}
\label{fig:ptrel}
\vspace{-1.cm}
\end{figure}

The measured cross-sections, for the process (2), as a function of \xgo \ 
and $p_T^{\rm rel}$ are shown in Figures \ref{fig:xgamma} and 
\ref{fig:ptrel} respectively, compared to {\sc Herwig} MC expectations. 
The ZEUS 
data points have statistical errors (inner bars) and statistical plus 
systematical errors added in quadrature (outer bars). The uncertainty due 
to the ZEUS calorimeter energy scale is shown as the shaded band. 
Figure \ref{fig:xgamma} shows the MC separately as 60\% direct 
photon processes (vertically 
hatched histogram) and 40\% resolved photon processes (diagonally 
hatched histogram) and the two combined (open histogram). In 
Figure \ref{fig:ptrel} the MC is shown separately as 83\% 
charm and light quark processes (diagonally hatched histogram) and 
17\%  beauty processes (horizontally hatched histogram) and again the two 
combined (open histogram). The percentages are the prediction 
from the MC, which requires a normalisation factor of 3.7 in order to describe 
the data. The {\sc Herwig} MC used has the default heavy quark 
masses; $m_c$~=~1.55~GeV and $m_b$~=~4.95~GeV and the CTEQ-4D \cite{cteq} and 
GRV-LO \cite{grv} structure functions for the proton and photon respectively.

%and the CTEQ-4D \cite{cteq} 
%structure function for the proton and GRV-LO \cite{grv} structure function 
%for the photon.

In Figure \ref{fig:xgamma}, we see a peak at high \xgo \ consistent 
with direct photon processes, but with a tail at low \xgo \ which cannot 
be explained by direct processes alone. The {\sc Herwig} MC predicts a 
significant component of resolved photon processes, which when added to 
the direct component shows good agreement in shape with the measured data. 
Fitting the ratio of direct and resolved to the data gives a resolved 
component of 35$\pm$6(\emph{stat})\%, in good agreement with the prediction 
and consistent with the aforementioned $D^*$ result \cite{dstar}. 

%It should be noted, however, that 
%the kinematic region in this analysis is not the same (although similar) as 
%that in \cite{dstar} and the cross-section presented here is not a property 
%of just the charm quark as expected in \cite{dstar}. 

In Figure \ref{fig:ptrel}, we see the cross-section peaked at low 
$p_T^{\rm rel}$, consistent with decays from predominantly charm quarks and
light mesons. At high $p_T^{\rm rel}$, we also see a significant 
cross-section consistent with decays predominantly from beauty quarks. 
The total MC sample shows good agreement with the measured data. Allowing 
the contributions of beauty and charm plus light flavours to vary, 
a percentage of $20 \pm 6\emph{(stat.)}^{+12}_{- \ 7}\emph{(syst.)}$ \% 
for beauty production ($\chi^2/$ d.o.f. $\sim 1.2$) was determined, in 
good agreement with the prediction. Using this value, we can then 
extract a beauty cross-section in our restricted kinematic region;

%\begin{eqnarray*}
%\vspace{-1.cm}
\[ \sigma_{b \bar{b}}^{\rm vis} (e^+ + p \rightarrow 
   {\rm dijet} +e^- + X) = 39 \pm 11 ^{+23}_{-16} \ {\rm pb}.\]
\begin{flushright}       {\rm (ZEUS \ Preliminary)}\end{flushright}
%\end{eqnarray*}

The cross-section has been compared to {\sc Herwig} MC predictions with 
the masses and structure functions specified previously. The measured value 
was found to lie a factor of about 4 above the MC prediction.

\section{Conclusions}

The first measurement of open beauty production from the ZEUS collaboration 
has been performed in dijet photoproduction events with an electron 
in the final state. The measured cross-section lies a factor of 4 above a 
LO MC prediction.
%and should therefore be confronted with other predictions. 
Although the measurement is interesting in itself, the 
opportunity to study the production mechanisms is the obvious goal.

\end{document}